\documentclass[prd,twocolumn,superscriptaddress,floatfix,amsmath,amssymb]{revtex4-1}
\usepackage{amsmath,amsfonts,amsthm,amssymb,slashed,graphicx}
\usepackage{xcolor}
\usepackage{graphicx}
\usepackage{epstopdf}
\usepackage{array,booktabs}
\usepackage{hyperref}
\usepackage{bbm} 
\usepackage{mathtools} 
\usepackage{tabulary}
\usepackage{multirow}
\usepackage{array,arydshln}
\hypersetup{linktoc = all}  
\hypersetup{pdfborderstyle={/S/U/W 0.5}}
\usepackage{tikz}
\usetikzlibrary{shapes.geometric}
\usetikzlibrary{intersections}
\usetikzlibrary{calc}
\newcounter{mycnt}

\newcommand{\tcr}{\textcolor{red}}

\begin{document}

\title{Super-Entropic Black Holes and the Kerr-CFT Correspondence}

\author{Musema Sinamuli}
\email{cmusema@perimeterinstitute.ca}
\affiliation{Department of Physics and Astronomy, University of Waterloo, Waterloo, Ontario N2L 3G1, Canada}
\affiliation{Perimeter Institute for Theoretical Physics, Waterloo, Ontario, N2L 2Y5, Canada}

\author{Robert B. Mann}
\email{rbmann@uwaterloo.ca}
\affiliation{Department of Physics and Astronomy, University of Waterloo, Waterloo, Ontario N2L 3G1, Canada}

\begin{abstract}
We demonstrate that Kerr-CFT duality can be extended to super-entropic black holes, which have non-compact horizons with finite area.    We demonstrate that this duality is robust insofar as  the ultra-spinning limit of a Kerr-AdS black hole (which yields the super-entropic class) commutes with the near-horizon limit (which yields the Kerr-CFT duality).   Consequently   the Bekeinstein-Hawking and the CFT entropies are equivalent.  We show that the duality holds
for both singly-spinning super-entropic black holes in 4 dimensions and for doubly-spinning super-entropic black holes of gauged supergravity in 5 dimensions.  In both cases we obtain not only
the expected left/right temperatures, but also temperatures associated with electric charge and with
a new thermodynamic parameter specific to super-entropic black holes.
\end{abstract}

\maketitle

\section{Introduction}

In the past few years  dualities between black holes and conformal field theories (CFT) have been of considerable interest, yielding new insights into our understanding of gravity.  One of the most prominent among them is the  ``Kerr/CFT correspondence" \cite{kerrcft}, which established a duality between the horizon of a Kerr black hole (bulk) and a $2d$ CFT. Many other intriguing results \cite{CFTdual}  have obtained since then. 

Recently a new class of rotating black hole solutions was obtained from the Kerr--Newman-AdS metrics \cite{Gnecchi:2013mja, Klemm:2014rda,superentropic}.   These black holes have been called {\it super-entropic} insofar as their entropy is larger than that expected from  the reverse isoperimetric inequality  conjecture \cite{blackhole}, which states that the thermodynamic volume of a black hole provides an upper bound on its entropy.  The basic idea of the procedure is to transform an azimuthal coordinate of a $d$-dimensional Kerr-AdS metric (written in rotating-at-infinity coordinates) and then take the limit as its associated rotation parameter approaches the AdS length.  The new azimuthal coordinate is then compactified, yielding a black hole whose horizon is topologically a $(d-2)$-sphere with two punctures.  These black holes form a new ultraspinning limit of the Kerr-AdS class of metrics, and possess a non-compact event horizon of finite area (and hence finite entropy), the first example of such objects in the literature to date.

The purpose of this paper is to study the Kerr-CFT correspondence for super-entropic black holes. 
It is not a-priori obvious that the correspondence exists, given the   
non-compactness of their event horizons; as we shall see some but not all super-entropic black holes
exhibit this correspondence.
 We will consider specifically the superentropic limit of singly-spinning Kerr-Newmann-AdS black holes in $d$ dimensions \cite{superentropic} as well as the ultraspinning $d=5$ black holes of minimal gauged supergravity \cite{general,ultraspin}.   Similar studies of Kerr-Newman-AdS black holes \cite{CFTdual} found that there exists a correspondence between the  extremal versions of these black holes and a $d=2$ (chiral half of a) CFT at its boundary.    A key feature of these metrics  is that in  their near-horizon (NH) limits, the resultant metrics acquire a new form that has an extended asymptotic symmmetry group (ASG)  whose generators depend strongly on the boundary conditions imposed. These boundary conditions have to be chosen in such a way that the charges associated with the diffeomorphisms preserving them satisfy a number of conditions such as finiteness, integrability and so on.
These  constraints ensure  the existence of the charges (generators of diffeomorphisms);  indeed, the consistency of the theory relies on them. Ignoring these constraints will certainly lead to ill-defined diffeomorphism generators from which no physics can be deduced.
 
 Moreover, the boundary conditions allow us to compute the central charges that stem from the Virasoro algebra satisfied by the charges associated with the diffeomorphisms preserving the boundary conditions. Once obtained, the central charges yield the CFT entropy $S_{CFT}$ via the Cardy formula, which is expected to be identical to the Bekenstein-Hawking $S$ entropy of the black hole, thereby establishing the correspondence. It is not {\it a-priori} obvious that super-entropic black holes admit such a correspondence, given their properties.  It is the purpose of this paper to demonstrate that   such a correspondence exists and indeed is robust.

 Our paper is organised as follows. In section two, we provide a brief review of the super-entropic black hole, highlighting its thermodynamic quantities and its relationship to the Smarr relation and the reverse isoperimetric inequality. 
In section three, we review the Kerr-CFT correspondence for  Kerr-AdS black holes, along with the choice of boundary conditions.  We then consider the Super-entropic-CFT correspondence, beginning with a Kerr-Newman-AdS black hole and taking both its super-entropic and near horizon limits in either order, and obtain the same results in either case.  In section five, we perform a similar analysis for super-entropic black holes of minimal gauged supergravity in $d=5$ \cite{general}. We conclude with some remarks on our results.
   
\section{Super-Entropic Black Holes}

We begin with a review of super-entropic black holes and their construction.

Let us start with the Kerr-Newman-AdS black hole metric \cite{ultraspin,superentropic}
\begin{eqnarray}
\label{KNAdS}
{ds}^2&=&-\frac{\Delta_{a}}{\Sigma_{a}}[d\bar t-a\sin^2\bar\theta \frac{d\bar\phi}{\Xi}]^2+\frac{\Sigma_{a}}{\Delta_{a}}d\bar r^2\nonumber\\
&+&\frac{\Sigma_{a}}{\mathcal{S}}d\bar\theta^2+\frac{\mathcal{S}\sin^2\bar\theta}{\Sigma_{a}}[ad\bar t-(\bar r^2+a^2)\frac{d\bar\phi}{\Xi}]^2\nonumber\\
\cal{A}&=&-\frac{q\bar r}{\Sigma_{a}}(d\bar t-a\sin^2\bar\theta \frac{d\bar\phi}{\Xi}) 
\end{eqnarray}
where 
\begin{eqnarray}
\Xi&=&1-\frac{a^2}{l^2},~~~\mathcal{S}=1-\frac{a^2}{l^2}\cos^2\bar\theta ,~~
\Sigma_{a}=\bar r^2+a^2\cos^2\bar\theta\nonumber\\
\omega_h&=&\frac{a}{a^2+r^2_+},
~~\Delta_{a}=(\bar r^2+a^2)(1+\frac{\bar r^2}{l^2})-2m\bar r+q^2\nonumber
\end{eqnarray}
Taking the ultraspinning limit $a\rightarrow l$ yields the super-entropic black hole, whose  metric is given by \cite{ultraspin,superentropic}
\begin{eqnarray}
\label{superentropic}
{ds}^2&=&-\frac{\Delta}{\Sigma}[d\bar t-l \sin^2\bar\theta d\bar\psi]^2+\frac{\Sigma}{\Delta}{d\bar r}^2\nonumber\\
&+&\frac{\Sigma}{\sin^2\bar\theta}{d\bar\theta}^2
+\frac{{\sin}^4\bar\theta}{\Sigma}[ld\bar t-(\bar r^2+l^2)d\bar\psi]^2\nonumber\\
\cal{A}=&-&\frac{q\bar r}{\Sigma}(d\bar t-l{\sin}^2\bar\theta d\bar\psi)
\end{eqnarray}
where
\begin{equation}
\Sigma=\bar r^2+l^2\cos^2\bar\theta ,~~~\Delta (\bar r)=(l+\frac{\bar r^2}{l})^2-2m\bar r+q^2
\end{equation}
provided the rescaled coordinate $\bar\psi$
\begin{equation}\label{barpsi}
\bar\psi=\frac{\bar\phi}{\Xi},~~~\Xi=1-\frac{a^2}{l^2}~~\mbox{with}~~a\rightarrow l  \quad .
\end{equation}
 is  taken to be periodic, identifying $\bar\psi \sim \bar\psi +\mu$. The quantity $\mu$   can be regarded as the chemical potential of the black hole \cite{superentropic} .

The function $\Delta (r)$ admits roots $r_\pm$ only when
\begin{equation}
m\geq m_{\ast}\equiv 2r_{\ast}(\frac{r^2_{\ast}}{l^2}+1)
\end{equation}
where
\begin{equation}
r^2_{\ast}\equiv \frac{l^2}{3}[-1+\sqrt{4+3\frac{q^2}{l^2}}]
\end{equation}
For an extremal black hole 
\begin{equation}
r_+=r_\ast~~~~~~  m=2r_{+}(\frac{r^2_{+}}{l^2}+1)
\end{equation}
and we note that extremality can occur even if $q=0$.

The super-entropic character of these black holes can be understood in the context of an {\em extended phase space} framework  \cite{Altamirano:2014tva},  where the cosmological constant is identified with the thermodynamic pressure according to
\begin{equation}
P=-\frac{\Lambda}{8\pi}
\end{equation}
in $d$ spacetime dimensions, with its conjugate quantity  treated as thermodynamic volume $V$.  The thermodynamic parameters of the super-entropic black hole are
\begin{eqnarray}
M&=&\frac{\mu m}{2\pi},~~J=Ml,~~Q=\frac{\mu q}{2\pi},~~\Phi=\frac{(q/l)x}{1+x^2},\nonumber\\
\omega_{h}&=&\frac{1}{l(1+x^2)}, ~~K=l\frac{(1-x^2)[(1+x^2)^2+q^2/l^2]}{8\pi x (1+x^2)},\label{TDparams}\\
A&=&2\mu l^2 (1+x^2),~~V=\frac{2}{3}l^3 \mu x (1+x^2)= \frac{l}{3}xA,\nonumber\\
T_{H}&=&\frac{1}{4\pi l x}\big[3 x^2-1-\frac{q^2/l^2}{1+x^2}\big],~~~ S =\frac{\mu}{2}l^2(1+x^2)=\frac{A}{4}\nonumber
\end{eqnarray}
 with $x=r_{+}/l$. The quantities $M, J, Q, \Phi, \omega_{h}, V, K,  T_{H}, A$ and $S $ are the mass, the angular momentum, the charge, the electric potential, the angular velocity at the horizon, the volume, the conjugate variable to the chemical potential $\mu$, the Hawking temperature, the area and the entropy of the black hole respectively. The first law of black hole thermodynamics is now extended to \cite{enthalpy}
\begin{equation}
\delta M= T\delta S+ \omega_h \delta J+\Phi\delta Q+V\delta P
\end{equation} 
with the quantities \eqref{TDparams} satisfying the  Smarr relation \cite{enthalpy}
\begin{equation}
\frac{d-3}{d-2}M=TS+\omega_h J+\frac{d-3}{d-2}\Phi Q-\frac{2}{d-2}VP
\end{equation}
with $d=4$.  The quantity $S$ in (\ref{TDparams}) 
 violates the {\it{Reverse Isoperimetric Inequality}}, which asserts \cite{blackhole}
\begin{equation}\label{RIPE}
{\cal{R}} \equiv {\bigg(\frac{(d-1)V}{\omega_{d-2}}\bigg)}^{\frac{1}{d-1}}\bigg(\frac{\omega_{d-2}}{A}\bigg)^{\frac{1}{d-2}} \geq 1
\end{equation}
where
\begin{equation}
\omega_{d}=\frac{\mu\pi^\frac{d-1}{2}}{\Gamma \big(\frac{d+1}{2}\big)} \quad .
\end{equation}
It is straightforward to show that for  (\ref{TDparams}) that
\begin{equation}
{\cal{R}}={\bigg(\frac{x^2}{1+x^2}\bigg)}^{\frac{1}{6}} < 1
\end{equation}
which clearly violates the inequality. The entropy of this class of black holes exceeds, for a given thermodynamic volume, the bound set by \eqref{RIPE}, and hence are called  super-entropic.

These singly spinning super-entropic black holes have extensions in any dimension, with  metric \cite{ultraspin}
\begin{eqnarray}
ds^2&=&-\frac{\Delta}{\rho^2}(d\bar t-l\sin^2\bar\theta d\bar \psi)^2+\frac{\rho^2}{\Delta}d\bar r^2+\frac{\rho^2}{\sin^2\bar\theta}d\bar\theta^2\nonumber\\
&+&\frac{\sin^4\bar\theta}{\rho^2}[ld\bar t-(\bar r^2+l^2)d\bar \psi]^2+\bar r^2\cos^2\bar\theta d\Omega^2_{d-4}
\end{eqnarray}
where
\begin{equation}
\Delta = \big(l+\frac{\bar r^2}{l}\big)^2-2m\bar r^{5-d},~~\rho^2=\bar r^2+l^2\cos^2\theta\nonumber
\end{equation}
and $d\Omega^2_d$ is the metric element on a $d$-dimensional sphere, where as before we identify
$\bar\psi \sim \bar\psi +\mu$. Writing $\varpi_d=\frac{\mu\pi^{\frac{d-1}{2}}}{\Gamma(\frac{d+1}{2})}$,
their thermodynamic parameters are
\begin{eqnarray}
M&=&\frac{\varpi_{d-2}}{8\pi}(d-2)m,~~\omega_h=\frac{l}{r^2_++l^2},\nonumber\\
T&=&\frac{1}{4\pi r_+l^2}[(d-5)l^2+r^2_+(d-1)],~~J=\frac{2}{d-2}Ml,\nonumber\\
S&=&\frac{\varpi_{d-2}}{4}(l^2+r^2_+)r^{d-4}_+=\frac{A}{4},~~V=\frac{r_+A}{d-1}
\end{eqnarray}
respectively denoting  the mass, the angular velocity at the horizon, the Hawking temperature, the angular momentum, the Bekenstein-Hawking entropy and the volume of the black holes respectively. 

While there exist horizons in any $d>5$ only when $m>0$ and for $d=5$ when $m>l^2/2$,
the extremal limit exists only in $d=4$. Specifically, extremal super-entropic black holes are those for which 
\begin{equation}
\label{restriction}
x^2=\frac{5-d}{d-1}
\end{equation}
whose only non-trivial solution is for $d=4$.  Henceforth we shall only consider this case.

\section{The Kerr-CFT limit} 

Here we review the Kerr-CFT correspondence for the Kerr-Newman-AdS black hole.
Returning to the metric (\ref{KNAdS}) in the extremal case and carrying out the near-horizon (NH) transformation
\begin{eqnarray}\label{kcftlim}
\bar{t}= r_{0}t/\epsilon  && \qquad  \bar{\theta} = \theta \nonumber\\
\bar{r}=  r_{+}+\epsilon r_{0} r  &&\qquad 
\bar{\phi} = \phi+\Xi\omega_{h}r_{0}t/\epsilon
\end{eqnarray}
where $\epsilon$ is a small parameter,  
yields a metric of the form 
\begin{eqnarray}
\label{generalform}
{ds}^2&=&\Gamma (\theta)[-r^2 dt^2+\frac{dr^2}{r^2}+\alpha (\theta)d\theta^2\nonumber\\
&+& \frac{\gamma (\theta)}{\Gamma (\theta)}(\frac{d\phi}{\Xi} + k r dt)^2]\nonumber\\
{\cal{A}}&=&f(\theta)(\frac{d\phi}{\Xi} +k r dt)
\end{eqnarray}
in the $\epsilon\to 0$ limit where  $k$ is  constant. The quantities $\Gamma, \alpha, \gamma, f$ are functions of the variable $\theta$ and can be computed from (\ref{KNAdS}). However  \eqref{generalform} is quite general: the metric of any $d=4$ stationary, axisymmetric extremal black hole, with a compact horizon section of non-toroidal topology, will have  a near-horizon limit of the form \eqref{generalform}, with $\Gamma, \alpha, \gamma, f$ 
taking specific values depending on the black hole studied \cite{NHsym,classification}.
Note that the new coordinates $(t,\phi, \theta, r)$ are dimensionless.  The quantity $r_{0}$ is a parameter with dimension of length whose value will be subsequently be fixed. 

The  metric \eqref{generalform} is invariant under the isometries \cite{CFTdual}
\begin{equation}
\bar{K_{1}}=\partial_{t},~~~ \bar{K_{2}}=t\partial_{t}-r\partial_{r}\nonumber
\end{equation}
\begin{equation}
\bar{K_{3}}=(\frac{1}{2r^2}+\frac{t^2}{2})\partial_{t}-tr\partial_{r}-\frac{k}{r}\partial_{\psi}, ~~~K_{1}=\partial_{\psi}
\end{equation}
which generate an $Sl(2,\mathbb{R})_{L}\times U(1)_{R}$ symmetry group.

 To study the corresponding CFTs, we will refer to the case of the Kerr-Newman- AdS black hole explored in \cite{CFTdual} and perform a similar analysis.    The metric function $\Delta (r)$ in (\ref{KNAdS}) 
has a root $r_+$ when
\begin{equation}
m\geq 2\texttt{r}_{\ast}\bigg(\frac{\texttt{r}^2_{\ast}}{l^2}+1\bigg)\nonumber
\end{equation}
with
\begin{equation}
\texttt{r}^2_{\ast}=\frac{l^2}{6}[-1-a^2/l^2+ \sqrt{1+14 a^2/l^2+a^4/l^4+ 12q^2/l^2}]\nonumber
\end{equation}
The thermodynamic quantities of the black hole (\ref{KNAdS})  are
\begin{eqnarray}
\label{thermoKNAdS}
M&=&m,~~J=Ma,~~Q=q,~~\Phi=\frac{qx}{l(a^2/l^2+x^2)},\nonumber\\
\omega_h&=&\frac{a/l^2}{a^2/l^2+x^2},~~A=\frac{4\pi}{\Xi}l^2(a^2/l^2+x^2),\nonumber\\
T_{H}&=&\frac{3x^2(a^2/l^2+x^2)+2x^2\Xi-(a^2/l^2+x^2)-q^2/l^2}{4\pi lx(a^2/l^2+x^2)},\nonumber\\
V&=&\frac{2\pi l^3}{3}\frac{(\frac{a^2}{l^2}+x^2)(2x^2+\frac{a^2}{l^2}-x^2\frac{a^2}{l^2})+\frac{q^2a^2}{l^4}}{\Xi^2 x}
\end{eqnarray}
respectively being its mass, the angular momentum, the charge, the electric potential, the angular velocity at the horizon, the horizon area, the Hawking temperature and  thermodynamic volume.   For an extremal black hole $T_H=0$ yielding  
\begin{equation}
r_+=\texttt{r}_*~~~~~~~m=2\texttt{r}_{*}(\frac{\texttt{r}^2_{*}}{l^2}+1)\nonumber
\end{equation} 
The Bekenstein-Hawking entropy as computed in \cite{ultraspin,superentropic} takes the value\begin{equation}\label{superS}
S=\frac{\pi}{\Xi}(a^2+r^2_{+})=\frac{A}{4} 
\end{equation}
noting that  $\phi$ is defined in the interval $[0, {2\pi})$.

We now aim to compute the CFT entropy of the Kerr-Newman-AdS black-hole (\ref{KNAdS}). To this end, we first take the near horizon limit \eqref{kcftlim}
 and obtain the metric  (\ref{generalform}), with 
\begin{eqnarray}
\Gamma (\theta)&=&\frac{\Sigma_{a_{+}}}{1+a^2/l^2+6x^2}\nonumber\\
\alpha (\theta)&=&\frac{1+a^2/l^2+6x^2}{\mathcal{S}}\nonumber\\
\gamma (\theta)&=&\frac{\mathcal{S}}{\Sigma_{a_{+}}}l^4(a^2/l^2+x^2)^2\sin^2\theta\nonumber\\
k&=&2\frac{(a/l)x}{(a^2/l^2+x^2)(1+a^2/l^2+6x^2)}\nonumber\\
r^2_{0}&=&l^2\frac{a^2/l^2+x^2}{1+a^2/l^2+6x^2}\nonumber\\
f(\theta)&=&q\frac{(1+x^2)}{2x}\frac{(x^2-a^2/l^2\cos^2\theta)}{x^2+a^2/l^2\cos^2\theta}
\label{NHKerr}
\end{eqnarray}
and $x=r_+/l$ as before.  The asymptotic symmetries of this metric   contain diffeomorphisms $\zeta$ such that \cite{CFTdual}
\begin{equation}
\delta_{\zeta}A_{\mu} = {\cal{L}}_{\zeta}A_{\mu} \qquad
\delta_{\zeta}g_{\mu\nu} = {\cal{L}}_{\zeta}g_{\mu\nu}
\end{equation}
as well as a $U(1)$ gauge transformation
\begin{equation}
\delta_{\Lambda}A=d\Lambda
\end{equation}
We will see shortly that the diffeomorphism and gauge transformations $(\zeta, \Lambda)$ obey an algebra and that their associated charges $Q_{\zeta,\Lambda}$ obey the same algebra up to a central charge term. The charge difference between two neighbouring metrics $g_{\mu\nu}$ and $g_{\mu\nu}+\delta g_{\mu\nu}$ is \cite{kerrcft,CFTdual}
\begin{equation}
\label{integral}
\delta Q_{\zeta,\Lambda}=\frac{1}{8\pi}\int (K_{\zeta}[h;g]+K_{\zeta,\Lambda}[h,a;g,A])
\end{equation}
with
\begin{equation}
a_{\mu}=\delta A_{\mu}, ~~~h_{\mu\nu}=\delta g_{\mu\nu}\nonumber
\end{equation}
\begin{eqnarray}
K_{\zeta}[h;g]&=&\frac{1}{4}\epsilon_{\alpha\beta\mu\nu}[\zeta^\nu D^\mu h-\zeta^\nu D_{\sigma}h^{\mu\sigma}+\zeta_{\sigma}D^{\nu}h^{\mu\sigma}\nonumber\\
&+&\frac{1}{2}hD^{\nu}\zeta^{\mu}-h^{\nu\sigma}D_{\sigma}\zeta^{\mu}\nonumber\\
&+&\frac{1}{2}h^{\sigma\nu}(D^{\mu}\zeta_{\sigma}+D_{\sigma}\zeta^{\mu})]dx^\alpha\times dx^\beta
\end{eqnarray}
and
\begin{eqnarray}
K_{\zeta,\Lambda}&=&\frac{1}{8}\epsilon_{\alpha\beta\mu\nu}[(-\frac{1}{2}hF^{\mu\nu}+2F^{\mu\sigma}{h_{\sigma}}^\nu\nonumber\\
&-&\delta F^{\mu\nu})(\zeta^\rho A_{\rho}+\Lambda)-F^{\mu\nu}\zeta^\sigma a_{\sigma}-2F^{\alpha\mu}\zeta^{\nu}a_{\alpha}\nonumber\\
&-&g^{\mu\sigma}g^{\nu\rho}a_{\sigma}({\cal{L}}_{\zeta}A_{\rho}+\partial_{\rho}\Lambda)]dx^\alpha\times dx^\beta
\end{eqnarray}
with~~ $\delta F^{\mu\nu}=g^{\mu\alpha}g^{\nu\beta}(\partial_\alpha a_\beta-\partial_\beta a_\alpha)$
 \vskip 5pt In order that the integral (\ref {integral}) be well define we must impose suitable boundary conditions which fulfil the conditions given in \cite{covariantthe,surfacecharge,brownhenneaux}. The choice of boundary conditions determines the asymptotic symmetry group (ASG), which is nothing but the allowed symmetries 
(those preserving the boundary conditions)
 modulo trivial ones (those for which the associated charges vanish). The idea is to make the boundary conditions as weak as we can whilst keeping the consistency of the theory. Consistency implies that the charges associated to the diffeomorphisms have to be finite (or may vanish). 
Choosing the same boundary conditions as in \cite{kerrcft} for the NH metric (\ref{generalform}) yields 
\[
h_{\mu\nu}\sim
\begin{bmatrix}
{\cal{O}}(r^2)& {\cal{O}}(1)& {\cal{O}}(1/r)& {\cal{O}}(1/r^2)\\
              & {\cal{O}}(1)& {\cal{O}}(1/r)& {\cal{O}}(1/r)\\
              &            & {\cal{O}}(1/r)& {\cal{O}}(1/r^2)\\
              &            &               & {\cal{O}}(1/r^3)
\end{bmatrix}
\]
and for the gauge field 
\begin{equation}
a_\mu\sim {\cal{O}}(r,1/r,1,1/r^2)
\end{equation}
all in the basis~ $(t,\phi,\theta, r)$.
 Making the choice 
 \begin{equation}
\Lambda=-f(\theta)\epsilon(\phi) \nonumber
\end{equation}
for the compensating gauge transformation \cite{CFTdual} 
satisfies the above boundary conditions.

The most general diffeomorphism preserving these boundary conditions is
\begin{eqnarray} 
\xi&=&[C+{\cal{O}}(1/r^3)]\partial_{t}+[-r\epsilon^{\prime}(\phi) +{\cal{O}}(1)]\partial_{r}\nonumber\\
&+&{\cal{O}}(1/r)\partial_{\theta}+[\epsilon+{\cal{O}}(1/r^2)]\partial_{\psi}
\end{eqnarray}
where $C$ is an arbitrary constant and $\epsilon(\phi)$ an arbitrary smooth function of $\phi$.  This includes the diffeomorphism 
\begin{equation}
\zeta=\epsilon\partial_{\psi}-r\epsilon^\prime\partial_{r}
\end{equation}
where $\epsilon^\prime = d\epsilon/d\phi$. This yields the Virasoro algebra 
\begin{equation}
i[\zeta_m,\zeta_n]=(m-n)\zeta_{m+n}~~~\mbox{with} ~~~\epsilon_n=-e^{-in\psi}\nonumber
\end{equation}
and
\begin{eqnarray}
[\Lambda_m,\Lambda_n]_\zeta&=&\zeta^{\mu}_{m}\partial_{\mu}\Lambda_n-\zeta^{\mu}_{n}\partial_{\mu}\Lambda_m\nonumber\\ 
i[\Lambda_m,\Lambda_n]_\zeta&=&(m-n)\Lambda_{m+n}\nonumber
\end{eqnarray}
which is the algebra of the ASG.

As noted above, the charges $Q_n$ associated with these diffeomorphisms and gauge transformations $(\zeta_n, \Lambda_n)$ satisfy a similar algebra
\begin{eqnarray}
&&i\{Q_m,Q_n\}= (m-n)Q_{m+n}+\frac{1}{8\pi}\int (K_{\zeta}[h;g]\nonumber\\
&&\qquad\qquad\qquad\qquad\qquad\qquad + K_{\zeta,\Lambda}[h,a;g,A])\nonumber\\
&&=(m-n)Q_{m+n} + \frac{c}{12}(m^3-\alpha m)\delta_{m+n,0}
\end{eqnarray}
the distinction being the central charge contribution,
where~$\alpha$ is a constant obtained after we parametrize $Q_n$. 
Proceeding  as in \cite{CFTdual} we therefore get a central charge
\begin{equation}
\label{centralc}
c
=\frac{3k}{\Xi}\int \sqrt{\Gamma (\theta)\gamma (\theta)\alpha (\theta)}d\theta
\end{equation}

 To determine the temperatures of the left and right-moving CFTs, we make use on one hand of the first law 
\begin{equation}
TdS=dM-(\omega_h dJ+{\Phi}dQ)
\end{equation}
and its extremality constraints
\begin{eqnarray}
\label{extremeconstraint}
T^{ex}dS&=&dM-(\omega^{ex}_h dJ+{\Phi}^{ex} dQ) 
= 0
\end{eqnarray} 
to obtain
\begin{eqnarray}
TdS&=&-[(\omega_h-\omega^{ex}_h) dJ+({\Phi}-{\Phi}^{ex}) dQ]
\end{eqnarray}
These variations can also been expressed as
\begin{equation}\label{dSvar}
dS=\frac{dJ}{T_L}+\frac{dQ}{T_e}
\end{equation}

We recall that  for a scalar field its expansion in eigenmodes of the energy and angular momentum is \cite{kerrcft}
\begin{equation}\label{scaleig}
\Phi=\sum_{E,J,s} \Phi_{E,J,s}~e^{-iE\bar t+iJ\bar\psi}f_s(r,\theta)
\end{equation}
for a Kerr-AdS black hole.
Near the horizon, the factor 
\begin{equation}
e^{-iE\bar t+iJ\bar\psi}\nonumber
\end{equation}
becomes 
\begin{eqnarray}
e^{-iE\bar t+iJ\bar\psi}&=&e^{-i(E-\omega^{ex}_{h}J)t r_0/\epsilon +iJ\psi} 
= e^{-in_Rt+in_L\psi}
\end{eqnarray}
upon using \eqref{kcftlim}, where
\begin{equation}
n_R = (E-\omega^{ex}_{h}J)r_0/\epsilon \qquad n_L = J
\end{equation}
For any system the density of states is $\rho=e^{S}$, with $S$ the entropy. Using this fact we extend the preceding expressions of $n_R$ and $n_L$ for a Kerr-Newman-AdS black hole to
\begin{equation}
n_R = (E-\omega^{ex}_{h}J-\Phi^{ex}Q)r_0/\epsilon \qquad n_L = J
\end{equation}
so that the density matrix in the energy and angular momentum eigenbasis has the Boltzmann weighting factor
\begin{equation}\label{Boltz}
e^{-(E-\omega_h J-\Phi Q)/T_H} =e^{-(n_R/T_R)-(n_L/T_L)-Q/T_e}
\end{equation}
and is a diagonal matrix when tracing over the modes inside the horizon. Comparing both sides of this equation
 yields the temperatures of the left and right-moving CFTs 
\begin{eqnarray}
T_L&=&-\frac{\partial T_H/\partial x}{\partial\omega_h/\partial x}\bigg |_{ex},~~~T_R=\frac {T_H r_0}{\epsilon}\bigg |_{ex}\nonumber\\
T_e&=&-\frac{\partial T_H/\partial x}{\partial\Phi/\partial x}\bigg |_{ex}
\end{eqnarray}
as well as the $T_e$ term in \eqref{Boltz}. For extremal black holes $T_H=0$ we see that 
$T_R\rightarrow 0$ and a
straightforward computation shows that the temperature of the left-moving CFT is~ $T_L=\frac{1}{2\pi k}$ and 
\begin{equation}
T_e=\frac{1}{2\pi q}\frac{[3(a^2/l^2+2x^2)+2\Xi-1](a^2/l^2+x^2)}{(x^2-a^2/l^2)}
\end{equation}
which is proportional to an inverse length $(\sim l^{-1})$.
 
The upshot of this exercise is that an extremal Kerr-Newmann-AdS black hole is dual to a $2d$ conformal field theory at its boundary with a mixed state whose  density matrix is expressed below.
 The Hartle-Hawking vacuum state is generalized around the Kerr-Newman-AdS black hole with a density matrix 
\begin{equation}
\rho=e^{-\frac{J}{T_L}-\frac{Q}{T_e}}
\end{equation}
Substituting our results into the CFT entropy from Cardy's formula, which states that the entropy of a unitary CFT at large $T$ or with $T\gg c$ satisfies
\begin{equation}\label{Cardy}
S_{CFT}=\frac{\pi^2}{3}c_L T_L
\end{equation}
yields 
\begin{equation}\label{SCFT1}
S_{CFT}=\frac{\pi}{\Xi}l^2(a^2/l^2+x^2)
\end{equation}
which is in agreement with the expression in \eqref{superS}.
Conditions for the applicability of \eqref{Cardy}
 have been given in \cite{kerrcft} in situations where $T\gg c$ does not hold. We shall see that a sufficient condition for the applicability of Cardy's formula in the super-entropic case is to set 
the electric charge $q$ to be large.
   
\section{Super-Entropic-CFT Correspondence} 

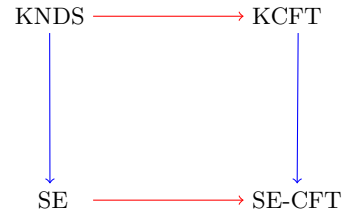
\begin{figure}
\centering
\begin{tikzpicture}
\path[draw] (0,0) node[below left](node A){$\phantom{K}$SE$\phantom{S}$}
            (2,0) node[below right](node B){SE-CFT}
            (2,2) node[above right](node C) {KCFT$\phantom{K}$}
            (0,2) node[above left](node D){KNDS};
\draw[red,->](node A)--(node B);
\draw[red,->](node D)--(node C);
\draw[blue,->](node C)--(node B);
\draw[blue,->](node D)--(node A);
\end{tikzpicture}
\caption{The above diagram indicates the various limits we are considering for the Kerr-Newman-AdS (KNDS) black hole: the Super-Entropic (SE) limit, the Kerr-CFT (KCFT) limit, and 
both limits together (SE-KCFT).  The horizontal arrows (in red) are the near horizon limit (NH) and vertical ones (blue) the ultraspinning limit. The resulting black hole is obtained by taking both limits one after another; we find that the same SE-KCFT limit results, indicating that the
square  commutes. }\label{fig:M1}
\end{figure}

In this section we establish that the ultraspinning super-entropic limit \eqref{superentropic}--\eqref{barpsi} commutes with the Kerr-CFT limit \eqref{kcftlim}, as shown in figure \ref{fig:M1}.

Let us first consider the lower path in figure \ref{fig:M1},  which consists  in starting with the Kerr-Newman-AdS black hole and taking the super-entropic limit, obtaining  (\ref{superentropic}). The next step is to take the near-horizon limit of the extremal version of (\ref{superentropic}). We obtain the metric (\ref{generalform}), but now with 
\begin{eqnarray}
\label{coeff1}
\Gamma (\theta)&=&\frac{l^2}{2}\frac{x^2+\cos^2\theta}{1+3x^2}\nonumber\\
\alpha (\theta)&=&\frac{2}{\sin^2\theta}(1+3x^2)\nonumber\\
\gamma (\theta)&=&l^2\sin^4\theta\frac{(1+x^2)^2}{x^2+\cos^2\theta}\nonumber\\
k&=&\frac{x}{(1+x^2)(1+3x^2)}\nonumber\\
r^2_{0}&=&\frac{l^2}{2}\frac{1+x^2}{1+3x^2}\nonumber\\
f(\theta)&=&q\frac{(1+x^2)}{x}\frac{x^2-\cos^2\theta}{x^2+\cos^2\theta}
\label{NHSE}
\end{eqnarray}
whose functions are also given by the $a\to l$ limit of \eqref{NHKerr}. 
The resulting metric likewise has the same topology as that of the metric
\eqref{superentropic}, with punctures  at the poles  \cite{ultraspin,superentropic}.

To compute the entropy of the corresponding CFT, the steps 
remain the same as described in the previous sections.  We impose the same boundary conditions and therefore the same diffeomorphisms and ASG. The central charge is now
\begin{eqnarray}
c&=&\frac{3k\mu}{2\pi}\int\sqrt{\Gamma (\theta)\gamma (\theta)\alpha (\theta)} d\theta\nonumber\\
&=&\frac{3k\mu}{\pi}l^2(1+x^2) = \frac{3x\mu l^2}{\pi (1+3x^2)}
\label{c-superent}
\end{eqnarray} 
using \eqref{coeff1}. We
remark that $c$ is very small when the electric charge $q$ becomes very large.

Considering the extremality constraint of the extended first law, we obtain
\begin{eqnarray}
T^{ex}dS&=&dM-(\omega^{ex}_h dJ+\Phi^{ex} dQ+ K^{ex}d\mu) = 0
\end{eqnarray} 
with $\mu$ the new thermodynamic variable whose conjugate is $K$ in \eqref{TDparams}. The first law then takes the form
\begin{equation}
TdS=-[(\omega_h-\omega^{ex}_h) dJ+(\Phi-\Phi^{ex}) dQ+ (K-K^{ex})d\mu]
\end{equation}
or alternatively 
\begin{equation}
dS=\frac{dJ}{T_L}+\frac{dQ}{T_e}+\frac{d\mu}{T_\mu}
\end{equation}

In this case the Boltzmann factor reads as
\begin{equation}
e^{-(E-\omega_hJ-\Phi Q-K\mu)/T_H}
\end{equation}
and we extend $n_R$ and $n_L$ as follows
\begin{equation}
n_R=(E-\omega^{ex}_{h}J-\Phi^{ex}Q-K^{ex}\mu)r_0/\epsilon ,~~~~n_L=J
\end{equation}
Then it takes the final form
\begin{equation}
e^{-n_R/T_R-n_L/T_L-Q/T_e-\mu/T_\mu}
\end{equation} 
Evaluating $n_R$ at the extremal limit, we find that it vanishes, unlike the situation for
the Kerr-CFT case. Hence $n_R$  can be interpreted as a quantity that measures a deviation from the extremal limit of a super-entropic black hole. 

It follows from the preceding relations that
\begin{eqnarray}
T_L&=&-\frac{\partial T_H/\partial r_+}{\partial\omega_h/\partial r_+}\bigg |_{ex}~~~~~T_R=\frac{r_0}{\epsilon}T_H\bigg |_{ex}\nonumber\\
T_e&=&-\frac{\partial T_H/\partial r_+}{\partial\Phi/\partial r_+}\bigg |_{ex}~~~~~T_\mu=-\frac{\partial T_H/\partial r_+}{\partial K/\partial r_+}\bigg |_{ex}
\end{eqnarray}
which are explicitly
\begin{eqnarray}
\label{lefttemp}
T_L&=&\frac{1}{2\pi k} = \frac{(1+x^2)(1+3x^2)}{2\pi x}~~~~~~~T_R=0\\
T_e&=&\frac{1}{\pi q}\frac{(3x^2+1)(x^2+1)}{(x^2-1)}~~~~~T_{\mu}=\frac{1}{2l^2}\frac{(1+3x^2)}{x^2}\nonumber
\end{eqnarray}
We remark that $T_{\mu}$ is a quantity inversely proportional to the square of a length $(\sim l^{-2})$. Furthermore, $T_L \gg c$  for sufficiently large $q$, justifying the use of   Cardy's formula \eqref{Cardy} at least in this regime. We therefore find upon insertion of  \eqref{c-superent}
into \eqref{Cardy} that
\begin{equation}\label{sentCardy}
S_{CFT}=\frac{\mu}{2}l^2(1+x^2)\nonumber
\end{equation}
for the extremal super-entropic black hole.

Turning now to the upper path in \ref{fig:M1}, we must take the limit
$a\to l$ in the metric \eqref{generalform} using the functions \eqref{NHKerr}.  This is straightforward and yields exactly the equations \eqref{NHSE}, where (as in \eqref{barpsi}) we must rescale $\phi\to \psi \Xi$ in \eqref{generalform}, identifying $\psi \sim \psi +\mu$ once the limit is taken.
The Hartle-Hawking vacuum density matrix becomes 
\begin{equation}
\rho=e^{-\frac{J}{T_L}-\frac{Q}{T_e}-\frac{\mu}{T_\mu}}
\end{equation}
with the temperatures identical to those in \eqref{lefttemp}.  Likewise, the   Cardy formula \eqref{Cardy} yields \eqref{sentCardy} for the CFT entropy using central charge and the left temperature obtained in (\ref{centralc}) and (\ref{lefttemp})respectively, thereby establishing the commutativity of the Kerr-CFT and super-entropic limits.

\section{Super-Entropic Black Holes of Gauged Supergravity}

The second set of black holes are black holes of minimal gauged $5d$ supergravity whose   metric is \cite{general}
\begin{eqnarray}
\label{5dminimal}
ds^2&=&d\gamma^2-\frac{2q\nu\omega}{\Sigma}+\frac{f\omega^2}{\Sigma^2}+\frac{\Sigma}{\Delta}d\bar r^2+\frac{\Sigma}{S}d\bar\theta^2\nonumber\\
A&=&\frac{\sqrt{3}q\omega}{\Sigma}
\end{eqnarray}
where
\begin{eqnarray}
d\gamma^2&=&-\frac{S\rho^2 d\bar t^2}{\Xi_a\Xi_b l^2}+\frac{\bar r^2+a^2}{\Xi_a}\sin^2\bar\theta d\bar\phi^2+\frac{\bar r^2+b^2}{\Xi_b}\cos^2\bar\theta d\bar{\tilde\psi}^2\nonumber\\
\nu&=&b\sin^2\bar\theta d\bar\phi+a\cos^2\bar\theta d\bar{\tilde\psi}\nonumber\\
\omega&=&\frac{Sd\bar t}{\Xi_a\Xi_b}-a\sin^2\bar\theta\frac{d\bar\phi}{\Xi_a}-b\cos^2\bar\theta\frac{d\bar{\tilde\psi}}{\Xi_b}\nonumber\\
S&=&\Xi_a\cos^2\bar\theta+\Xi_b\sin^2\bar\theta\nonumber\\
\Delta&=&\frac{(\bar r^2+a^2)(\bar r^2+b^2)\rho^2/l^2+q^2+2abq}{\bar r^2}-2m\nonumber\\
\Sigma&=&\bar r^2+a^2\cos^2\bar\theta+b^2\sin^2\bar\theta,~~\rho^2=\bar r^2+l^2\nonumber\\
\Xi_a&=&1-\frac{a^2}{l^2},~~\Xi_b=1-\frac{b^2}{l^2}\nonumber\\
f&=&(2m+\frac{2abq}{l^2})\Sigma-q^2
\end{eqnarray}
Considering coordinates that rotate at infinity
\begin{equation}
\bar\phi=\bar\phi_R+\frac{a}{l^2}t,~~~\bar{\tilde\psi}=\bar\psi_R+\frac{b}{l^2}t
\end{equation}
we rewrite the metric in the more suitable form \cite{extremal}
\begin{equation}
\label{metric}
ds^2=-e^0 e^0+\sum_{i=1}^{4}e^i e^i
\end{equation}
with 
\begin{eqnarray}
\label{vielbein}
e^0&=&\sqrt{\frac{\Delta}{\bar r^2+y^2}}\omega,~~e^1=\sqrt{\frac{\bar r^2+y^2}{\Delta}}d\bar r^2,\\ e^2&=&\sqrt{\frac{Y}{\bar r^2+y^2}}(dt^{'}-\bar r^2d\psi_1),~~e^3=\sqrt{\frac{\bar r^2+y^2}{Y}}dy,\nonumber\\
e^4&=&\frac{ab}{\bar ry}\bigg(dt^{'}+(y^2-\bar r^2)d\psi_1-\bar r^2y^2d\psi_2+\frac{qy^2}{ab(\bar r^2+y^2)}\varpi\bigg)\nonumber
\end{eqnarray}
and
\begin{eqnarray}
\label{vielbein2}
Y&=&-\frac{(1+y^2/l^2)(a^2-y^2)(b^2-y^2)}{ y^2},~~\varpi=dt^{'}+y^2d\psi_1\nonumber\\
t^{'}&=&\bar t-(a^2+b^2)\psi_1-a^2b^2\psi_2\nonumber\\
\psi_1&=&\frac{a}{a^2-b^2}\frac{\bar\phi_R}{\Xi_a}+\frac{b}{b^2-a^2}\frac{\bar\psi_R}{\Xi_b}\nonumber\\
\psi_2&=&\frac{1}{a(b^2-a^2)}\frac{\bar\phi_R}{\Xi_a}+\frac{1}{b(a^2-b^2)}\frac{\bar\psi_R}{\Xi_b}\nonumber\\
y^2&=&a^2\cos^2\bar\theta+b^2\sin^2\bar\theta
\end{eqnarray}
Let us introduce new notations $\bar\varphi=\bar\phi_R/\Xi_a$ and $\bar\psi=\bar\psi_R/\Xi_b$. The thermodynamics quantities for this black hole are
\begin{eqnarray}
\omega_\varphi&=&\frac{a(r^2_++b^2)+bq}{(r^2_++a^2)(r^2_++b^2)+abq}\nonumber\\
\omega_{\psi}&=&\frac{b(r^2_++a^2)+bq}{(r^2_++b^2)(r^2_++b^2)+abq}\nonumber\\
T_H&=&\frac{r^4_+[1+(2r^2_++a^2+b^2)/l^2]-(q+ab)^2}{2\pi r_+[(r^2_++a^2)(r^2_++b^2)+abq]}\nonumber\\
S&=&\frac{\pi^2[(r^2_++a^2)(r^2_++b^2)+abq]}{2\Xi_a\Xi_br_+}
\end{eqnarray}
which are the angular velocity in the direction $\bar\varphi$ at the horizon, the angular velocity in the direction $\bar\psi^{'}$ at the horizon, the Hawking temperature and the entropy of the black hole respectively.
In the case of an extremal black hole $r_+$ solves 
\begin{equation}
\label{extremegauged}
r^{4}_{+}[1+(2r^{2}_{+}+a^2+b^2)/l^2]-(q+ab)^2=0
\end{equation}
This equation admits a positive root that grows as long as the charge $q$ gets larger.

Let us consider first the upper path 
as in \ref{fig:M1}.   Assuming extremality via \eqref{extremegauged}, upon taking the NH limit 
\begin{eqnarray}
\label{NHsugra}
\bar t&=&tr_0/\epsilon ~~~~~\bar\theta=\theta ~~~~
\bar r=r_++r_0r\epsilon\nonumber\\
\bar\varphi&=&\varphi+\omega_\varphi tr_0/\epsilon ~~~~~
\bar\psi=\psi+\omega_{\psi}tr_0/\epsilon
\label{NHsugra}
\end{eqnarray}
 the vielbeins (\ref{vielbein}) take the form 
\begin{eqnarray}
\label{vielb}
e^0&=&\sqrt{\Gamma(\theta)}rdt,~~e^1=\sqrt{\Gamma(\theta)}\frac{dr}{r},\nonumber\\
e^2&=&\alpha_1e_1+\alpha_2e_2,~~
e^3=\sqrt{\Gamma(\theta)\alpha(\theta)}d\theta\nonumber\\
e^4&=&\beta_1 e_1+\beta_2 e_2
\end{eqnarray}
where
\begin{eqnarray}
\label{function1}
\Gamma (\theta)&=&\frac{l^2}{4}\frac{\Sigma_+}{3r^2_++l^2+a^2+b^2}\\
\alpha (\theta)&=&\frac{\Sigma_+}{(\Xi_a\cos^2\theta+\Xi_b\sin^2\theta)\Gamma (\theta)}\nonumber\\
e_1&=&d\varphi+k_1rdt,~~e_2=d\psi+k_2rdt\nonumber\\
r^2_0&=&\frac{l^2[(r^2_++a^2)(r^2_++b^2)+abq]}{4 r^2_+(3r^2_++l^2+a^2+b^2)}\nonumber\\
k_1&=&\frac{l^2[[a(r^2_++b^2)+bq](r^2_++b^2)+ bqr^2_+]}{2r_+[(r^2_++a^2)(r^2_++b^2)+abq][3r^2_++l^2+a^2+b^2]}\nonumber\\
k_2&=&\frac{l^2[[b(r^2_++a^2)+aq](r^2_++a^2)+ aqr^2_+]}{2r_+[(r^2_++a^2)(r^2_++b^2)+abq][3r^2_++l^2+a^2+b^2]}\nonumber
\end{eqnarray}
and
\begin{eqnarray}
\label{function2}
\alpha_1&=&a\sqrt{\frac{Y}{r^2_++y^2}}\frac{r^2_++a^2}{a^2-b^2}\nonumber\\
\alpha_2&=&b\sqrt{\frac{Y}{r^2_++y^2}}\frac{r^2_++b^2}{b^2-a^2}\nonumber\\
\beta_1&=&\frac{(a^2-y^2)[aqy^2+b(r^2_++a^2)(r^2_++y^2)]}{r_+y(a^2-b^2)(r^2_++y^2)}\nonumber\\
\beta_2&=&\frac{(b^2-y^2)[bqy^2+a(r^2_++b^2)(r^2_++y^2)]}{r_+y(b^2-a^2)(r^2_++y^2)}
\end{eqnarray}

 The computation of the central charge requires the choice of boundary conditions 
\begin{equation}
h_{\mu\nu}\sim
\begin{bmatrix}
{\cal{O}}(r^2)&{\cal{O}}(1)&{\cal{O}}(r^2)&{\cal{O}}(1/r)&{\cal{O}}(1/r^2)\\
      &{\cal{O}}(1)&{\cal{O}}(1)&{\cal{O}}(1/r)&{\cal{O}}(1/r)\\
              &   &{\cal{O}}(1)&{\cal{O}}(1/r)&{\cal{O}}(1/r)\\
              &            &    &{\cal{O}}(1/r)&{\cal{O}}(1/r^2)\\
              &            &               &   &{\cal{O}}(1/r^3)
\end{bmatrix}
\label{BC-gauged}
\end{equation} 
in the basis $(t,\varphi,\psi,\theta,r)$. 
The diffeomorphism that preserves these boundary conditions is
\begin{eqnarray}
\xi&=&[C+{\cal{O}}(1/r^3)]\partial_{t}+[-r(\epsilon^{\prime}(\varphi) +\epsilon^{\prime}(\psi)) +{\cal{O}}(1)]\partial_{r}\\
&+&{\cal{O}}(1/r)\partial_{\theta}+[\epsilon(\varphi)+{\cal{O}}(1/r^2)]\partial_{\varphi}+[\epsilon(\psi)+{\cal{O}}(1/r^2)]\partial_{\psi}\nonumber
\end{eqnarray}
It clearly contains the diffeomorphisms
\begin{eqnarray}
\zeta_\varphi&=&\epsilon(\varphi)\partial_\varphi-r\epsilon^{\prime}(\varphi) \partial_{r}\nonumber\\
\zeta_\psi&=&\epsilon(\psi)\partial_{\psi}-r\epsilon^{\prime}(\psi)\partial_{r}
\end{eqnarray}
When following the same steps than the previous cases the diffeormorphism $\zeta_\varphi$ gives rise to central charge
\begin{eqnarray}
\label{centralcharge1}
c_\varphi&=&\frac{3 k_1}{8\pi}\int\sqrt{\Gamma(\theta)\alpha(\theta)(\alpha_2\beta_1-\alpha_1\beta_2)^2}d\theta d\varphi d\psi\nonumber\\
&=&\frac{3\pi k_1[(r^2_++a^2)(r^2_++b^2)+abq]}{2r_+\Xi_a\Xi_b}
\end{eqnarray}
Similarly for the diffeomorphism $\zeta_\psi$ 
\begin{eqnarray}
\label{centralcharge2}
c_\psi&=&\frac{3k_2}{8\pi}\int\sqrt{\Gamma(\theta)\alpha(\theta)(\alpha_2\beta_1-\alpha_1\beta_2)^2}d\theta d\varphi d\psi\nonumber\\
&=&\frac{3\pi k_2[(r^2_++a^2)(r^2_++b^2)+abq]}{2r_+\Xi_a\Xi_b}
\end{eqnarray}
The first law of thermodynamics and the extremality
constraint \eqref{extremegauged} give
\begin{equation}
TdS=-[(\omega_\varphi-\omega^{ex}_\varphi)dJ_\varphi+
(\omega_{\psi}-\omega^{ex}_{\psi})dJ_\psi+
(\Phi-\Phi^{ex})dQ]
\end{equation}
and can be rewritten as
\begin{equation}
dS=\frac{dJ_\varphi}{T_\varphi}+\frac{dJ_\psi}{T_\psi}+\frac{dQ}{T_e}
\end{equation}
The Boltzmann factor for black holes of gauged supergravity reads as
\begin{equation}
e^{-(E-\omega_\varphi J_\varphi-\omega_{\psi}J_\psi-\Phi Q)/T_H}=e^{-n_R/T_R-n_\varphi/T_\varphi-n_\psi/T_\psi-Q/T_e}
\end{equation}
where 
\begin{equation}
n_R=(E-\omega^{ex}_\varphi J_\varphi-\omega^{ex}_{\psi}J_\psi-\Phi^{ex} Q)r_0/\epsilon~~~~n_{\varphi,\psi}=J_{\varphi,\psi}
\end{equation}
We then get the temperatures of the left and right-moving CFT's as well as the quantity $T_e$ associated to the electric charge
\begin{eqnarray}
T_\varphi&\equiv&-\frac{\partial T_H/\partial r_+}{\partial\omega_\varphi/\partial r_+}\bigg |_{ex}=\frac{1}{2k_\varphi},~~T_\psi\equiv-\frac{\partial T_H/\partial r_+}{\partial\omega_{\psi}/\partial r_+}\bigg |_{ex}=\frac{1}{2k_\psi}\nonumber\\
T_R&\equiv&\frac{T_H r_0}{\epsilon}\bigg|_{ex}=0,~~~~~ T_e\equiv-\frac{\partial T_H/\partial r_+}{\partial\Phi/\partial r_+}\bigg |_{ex}
\end{eqnarray}
The applicability of  the Cardy formula \eqref{Cardy} again
requires large $q$ and small rotation parameter compared to the AdS raduis $a\ll l$. Thus, the CFT entropy is 
\begin{equation}
S_{CFT}= \frac{\pi^2}{3}c_\varphi T_\varphi+\frac{\pi^2}{3}c_\psi T_\psi 
=\frac{[(r^2_+ +a^2)(r^2_+ +b^2)+abq]}{2r_+\Xi_a\Xi_b}
\end{equation}
for the extremal black hole \eqref{metric}.

The super-entropic limit of \eqref{metric} can only be taken in one azimuthal direction \cite{ultraspin}.  Without loss of generality, we  choose this be the $\phi$-direction, setting $a\to l$ and requiring
the new coordinate $\varphi$ to be periodic with period $\mu$.   Replacing $\bar\phi_R/\Xi_a$
with $\varphi$, the vielbeins and associated parameters are obtained by $a=l$ in (\ref{vielbein})  and (\ref{vielbein2}) respectively. The thermodynamic quantities of this super-entropic black hole are \cite{ultraspin}
\begin{eqnarray}
M&=&\frac{\mu}{8}\frac{(m+bq/l)(2+\Xi_b)}{\Xi_b},~~J_\varphi=\frac{\mu}{4}\frac{lm+bq}{\Xi_b},\nonumber\\
J_{\psi}&=&\frac{\mu}{8}\frac{2bm+q(b^2+l^2)/l}{\Xi_b},\nonumber\\
\omega_\varphi&=&\frac{l(r^2_++b^2)+bq}{(r^2_++l^2)(r^2_++b^2)+lbq},\nonumber\\
\omega_{\psi}&=&\frac{b(r^2_++l^2)+lq}{(r^2_++l^2)(r^2_++b^2)+lbq},\nonumber\\
T_H&=&\frac{r^4_+[2+(2r^2_++b^2)/l^2]-(q+bl)^2}{2\pi r_+[(r^2_++l^2)(r^2_++b^2)+lbq]},\nonumber\\
S&=&\frac{\mu\pi[(r^2_++l^2)(r^2_++b^2)+lbq]}{4r_+\Xi_b}=\frac{A}{4},\nonumber\\
\Phi&=&\frac{\sqrt{3}qr^2_+}{(r^2_++l^2)(r^2_++b^2)+lbq},~~Q=\frac{\mu\sqrt{3}q}{8\Xi_b}
\end{eqnarray}
and are the mass, the angular momentum in the direction $\varphi$, the angular momentum in the direction $\psi$, the angular velocity in the direction $\varphi$ at the horizon, the angular velocity in the direction $\psi$ at the horizon, the Hawking temperature, the Bekeinstein-Hawking entropy, the electric potential and the electric charge of the black holes respectively.

Upon taking the NH limit \eqref{NHsugra} for the extremal case
 \begin{equation}
\label{extremegauged-sent}
r^{4}_{+}[1+(2r^{2}_{+}+l^2+b^2)/l^2]-(q+lb)^2=0
\end{equation} 
 we obtain (\ref{vielb},\ref{function1},\ref{function2}) but with $a=l$.   This is the same
 result as that would be obtained if the super-entropic limit of  (\ref{vielb}) were taken.

After fixing the same boundary conditions \eqref{BC-gauged} we find
\begin{equation}
\label{cc1}
c_\varphi=\frac{3k_1\mu[(r^2_++l^2)(r^2_++b^2)+lbq]}{4r_+\Xi_b}
\end{equation}
and
\begin{equation}
\label{cc2}
c_\psi=\frac{3k_2\mu[(r^2_++l^2)(r^2_++b^2)+lbq]}{4r_+\Xi_b}
\end{equation}
 the central charges, a result valid for both the upper and lower paths of Fig. \ref{fig:M1}.
 
For either path the first law of thermodynamics and its extremality
constraint  are 
\begin{eqnarray}
TdS&=&-[(\omega_\varphi-\omega^{ex}_\varphi)dJ_\varphi+
(\omega_{\psi}-\omega^{ex}_{\psi})dJ_\psi\nonumber\\
&+&(\Phi-\Phi^{ex})dQ + (K-K^{ex})d\mu]
\end{eqnarray}
which we write as
\begin{equation}
dS=\frac{dJ_\varphi}{T_\varphi}+\frac{dJ_\psi}{T_\psi}+\frac{dQ}{T_e}+\frac{\mu}{T_\mu}
\end{equation}
and the Boltzmann factor is
\begin{equation}
e^{-\frac{1}{T_H}(E-\omega_\varphi J_\varphi-\omega_{\psi}J_\psi-\Phi Q-K\mu)}=e^{-\frac{n_R}{T_R}-\frac{n_\varphi}{T_\varphi}-\frac{n_\psi}{T_\psi}-\frac{Q}{T_e}-\frac{\mu}{T_\mu}}
\end{equation}
where 
\begin{eqnarray}
n_R&=&(E-\omega^{ex}_\varphi J_\varphi-\omega^{ex}_{\psi}J_\psi-\Phi^{ex}Q-K^{ex}\mu)\frac{r_0}{\epsilon}\nonumber\\
n_{\varphi,\psi}&=&J_{\varphi,\psi}
\label{n-gauged}
\end{eqnarray}
The temperatures of the two left and one right-moving CFT's and the quantities $T_e$ and $T_\mu$ respectively associated with the electric charge and the chemical potential are
\begin{eqnarray}
T_\varphi&\equiv&-\frac{\partial T_H/\partial r_+}{\partial\omega_\varphi/\partial r_+}\bigg |_{ex}=\frac{1}{2k_\varphi},~~T_\psi\equiv-\frac{\partial T_H/\partial r_+}{\partial\omega_{\psi^{'}}/\partial r_+}\bigg |_{ex}=\frac{1}{2k_\psi}\nonumber\\
T_R&\equiv&\frac{T_H r_0}{\epsilon}\bigg|_{ex}=0,~~~~T_e\equiv-\frac{\partial T_H/\partial r_+}{\partial\Phi/\partial r_+}\bigg |_{ex}\nonumber\\
T_{\mu}&\equiv&-\frac{\partial T_H/\partial r_+}{\partial K/\partial r_+}\bigg |_{ex}
\end{eqnarray}
upon comparing the two equations in \eqref{n-gauged}. 

Finally, we find that the CFT entropy is
\begin{equation}
S_{CFT}=\frac{\mu\pi[(r^2_++l^2)(r^2_++b^2)+lbq]}{4r_+\Xi_b}
\end{equation}  
for both paths.  As before, a sufficient condition for the Cardy formula \eqref{Cardy} is that the electric charge $q$ is sufficiently large, thereby ensuring that $T_\varphi \gg c$ and $T_\psi \gg c$.

\section{Conclusion}

We have demonstrated the super-entropic black holes, despite the non-compactness of their horizons, have well-defined Kerr-CFT correspondence limits.  These limits are robust: the CFT limit of a super-entropic black hole yields the same results as the super-entropic limit of an extremal 
near-horizon Kerr-AdS metric. 
Indeed, we verified that we always end up with the same outcome depending on whether we follow either the upper or lower path in Fig. \ref{fig:M1}.

A remarkable feature of super-entropic black holes is that the new variable $\mu$, interpreted as the chemical potential and obtained from compactification of the azimuthal direction, not only enters into the Cardy formula to yield an entropy for the CFT, but also yields a new 
 quantity $T_{\mu}$ that appears in the Hartle-Hawking density matrix. The latter scales as the inverse of the square of a length $(\sim l^{-2})$ in $4d$ or more generally as $(\sim l^{2-d})$ in all dimensions $d$.
 
We note that the Kerr-CFT correspondence only applies for singly-spinning super-entropic black holes in $d=4$, since the extremality condition does not hold for $d>4$. The $d=5$ gauged supergravity super-entropic black holes, however, do exhibit the correspondence. In both cases, 
a sufficient condition for the applicability of the Cardy formula \eqref{Cardy} is that 
the electric charge of such black holes is taken to be large.  This is in contrast to both
 Kerr-Newman-AdS and gauged supergravity black holes in which both 
large electric charges and  small rotation parameters compared to the AdS radii $(a\ll l)$, are required.  
  
 We expect that the Kerr-CFT correspondence for the recently obtained
 multiply spinning super-entropic black holes 
\cite{ultraspin} can be established using arguments similar to the ones that we have presented.

\section*{Acknowledgements} 
This work was supported in part by the Natural Sciences and Engineering Research Council of Canada.

\end{document}